\documentclass{article}

\usepackage{hyperref}
\usepackage{color}
\usepackage{graphicx}
\usepackage{amstext,amsmath,amssymb,amsfonts,bbm,
amsthm}
\usepackage{fancyhdr}
\usepackage{subfigure}
\usepackage{float}

\begin{document}


\title{Euler-Heisenberg waves propagating in a magnetic background}

\author{Elda Guzman-Herrera\footnote{eguzman@fis.cinvestav.mx} and 
Nora Breton\footnote{nora@fis.cinvestav.mx} \vspace{1cm}\\
Departamento de F\'{i}sica \\
Centro de Investigaci\'{o}n y de Estudios Avanzados del I. P. N. \\ 
Apdo. 14-740, CDMX, M\'{e}xico}

\maketitle

\begin{abstract}
We derive the Euler-Heisenberg solutions that describe electromagnetic waves propagating through very intense uniform magnetic or electric background, with the effective metric approach. We first explore the case of a magnetic background: 
as a result of the interaction between the wave and the background there is birefringence and  a longitudinal electric field component arises.  The two phase velocities depend on the intensity of the external magnetic field and on the polarization of the wave;  phase velocities can be slowed down up to the order of hundred thousandths for fields $B/B_{\rm cr} << 1$. The analogous study is done when the wave propagates through a uniform electric field.
We then consider the situation when the background is in movement by means of a Lorentz boost, modeling then a magnetized flowing medium. We determined how this motion affects the speed of propagation of the electromagnetic wave, in this case the phase velocities  depend on both the magnetic background and the direction and velocity of the boost.
\end{abstract}
\vspace{1cm}

PACS: 12.20.Ds, 11.10.Wx, 41.20.Jb

\newpage

\section{Introduction}

In the presence of intense electromagnetic fields quantum electrodynamics (QED) predicts that vacuum has properties of a material medium as a consequence of the electromagnetic field  self-interactions. 
When the electromagnetic field strengths approach the critical fields
$E_{cr} \approx m_e^2 c^3/(e \hbar) \approx 10^{18}$ Volt/m or 
$B_{cr} \approx 10^{9}$ Tesla the impact of the external field on quantum processes  becomes significant.

The Euler–Heisenberg (EH) Lagrangian  was  derived from QED principles by W. Heisenberg and H. Euler in 1936 \cite{EH}. For a nice discussion on the history of the Euler-Heisenberg approach see \cite{Dunne2012} and a pedagogical review can be found in \cite{Dunne2004}. 
By treating the vacuum as a medium, EH effective action predicts nonlinear light interaction processes since
it takes into account vacuum polarization to one loop, and is valid for electromagnetic fields that change slowly compared to the inverse electron mass. 
The EH Lagrangian ${\mathcal{L}}_{\rm EH} (F,G) $ depends in nonlinear way on the two Lorentz and gauge invariants of the Faraday tensor $F_{\mu \lambda}$, $F=F^{\mu \lambda}F_{\mu \lambda}= 2(B^2-E^2)$ and $G=F^{*\mu\lambda}F_{\mu\lambda}= - 4 \vec{B} \cdot \vec{E}$, with the dual field $F^{*\mu\lambda}= \frac{1}{2} \epsilon^{\mu \lambda \alpha \beta} F_{\alpha \beta}$, 

\begin{equation}
{\mathcal{L}}_{\rm EH} (F,G) = -\frac{F}{4} -\frac{1}{8 \pi^2} \int_{0}^{\infty}{ e^{-m^2 s} \left[ (es)^2 \frac{{\rm Re} [\cosh(es \sqrt{2(F+iG)})]}{{\rm Im}[\cosh(es \sqrt{2(F+iG)})]}- \frac{2}{3}(es)^2 F -1 \right] \frac{ ds}{s^3}}.    
\label{E-HLagrangian_complete}
\end{equation}

From this Lagrangian new nonlinear interactions can be derived, which do not occur in the tree level Maxwell action; among them are light-light interaction and pair production from vacuum excited by an electromagnetic field. It has been thoroughly investigated, and higher loop contributions in strong fields have been calculated as well, see for instance \cite{GiesKarbstein2017}, \cite{Karbstein2019}.

The Lagrangian (\ref{E-HLagrangian_complete}) can be expanded into an asymptotic series \cite{Adler1970}, \cite{Ritus1972} whose first terms, of the order $\alpha^2$, are 

\begin{equation}
\mathcal{L}_{\rm EK} (F,G) = -\frac{F}{4}+ \frac{\mu}{4}\left(F^{2}+\frac{7}{4}G^{2} \right),
\label{E-HLagrangian}
\end{equation}
where $\mu$ is the parameter of the EH theory that in terms of the fine structure constant, $\alpha$ ($c=1,\quad h=1$), is
\begin{equation}
\mu= \frac{2 \alpha^2}{45m_e^4};
\label{b}
\end{equation}
that in terms of the critical fields, is of the order $\mu \sim \alpha/ B_{cr}^2$. 
Actually (\ref{E-HLagrangian})  is the Euler-Kockel (EK) Lagrangian. H. Euler and B. Kockel, two Heisenberg's students, 
investigated on QED vacuum polarizations  in the constant background limit, obtaining the leading nonlinear correction in powers of the field strengths, presenting the Lagrangian (\ref{E-HLagrangian}) in 1935 \cite{EK1935}.
To describe the propagation of photons in an external field in QED only this first correction is needed, i.e. the term $2 \alpha^2 (F^2+7 G^2/4)/(45 m_e^4)$.
The use of the first terms of this expansion is justified if the dimensionless expansion parameter $4 \pi \alpha \hbar^3 {\lvert F \rvert}^2/(m_e^4 c^5)$ is much smaller than unity \cite{Bialynicka1970}. This is indeed the case even for strong magnetic fields, for instance the magnetic fields in neutron stars that may be as large as $10^{12}$ Gauss \cite{Gold1968}, then processes like photon splitting and pair conversion are expected to occur in the vicinity of neutron stars \cite{Baring2008}.

We also note related work by Born and Infeld (1934) \cite{BI1934}, who obtained similar nonlinear corrections to Maxwell theory but from a classical perspective. See \cite{Kruglov2010} for a generalized Born-Infeld electrodynamics. The linear electromagnetic Maxwell theory is recovered if $\mu=0$, $\mathcal{L}_{\rm Maxwell} (F) = -{F}/{4}$.

Vacuum polarization effects \cite{Schwinger1951}, \cite{DittrichGies1998}, \cite{DittrichGies2000} have been studied since the
establishment of these Lagrangians and experimental
efforts are currently in progress for measuring some of these nonlinear  effects, we mention just a few of them:
Light by light interactions can be studied using heavy-ion collisions; the electromagnetic (EM) field strengths  produced, for example by a Pb nucleus would be up to $10^{25}\rm{Vm^{-1}}$, those intense EM fields can be treated as a beam of quasi-real photons, and it has been measured light by light scattering in $Pb+ Pb$ collisions at the Large Hadron Collider \cite{Atlas2017}. Other experimental evidences include the measurement of photon splitting in strong magnetic fields \cite{Akhmadaliev2002} and the search for vacuum polarization with laser beams crossing magnetic fields  or the detection of vacuum birefringence with intense laser pulses \cite{Luiten2004}. There is also the detection of QED vacuum nonlinearities using  waveguides \cite{Brodin2001}. Vacuum pair production, known as the Sauter-Schwinger effect \cite{Schwinger1951},  was a prediction in the 1936 EH paper however the necessary electric field strengths are not reached yet, since the corresponding critical laser intensity is about $I_{\rm cr}= 4.3 \times 10^{29}  {\rm W/ cm}^2$ \cite{Gelis2016}.

The phase velocity of an electromagnetic wave traveling through intense EM fields will be altered due to vacuum polarization. It is also well known the emergence of a longitudinal field component, as well as the decrease of the phase velocity \cite{Adler1970}, \cite{Batalin1971}, \cite{Bialynicka1970}. Another effect that arises in strong magnetic backgrounds is the birefringence \cite{Brezin1971}.
In order to attempt the detection of nonlinear electromagnetic effects 
it is crucial to determine the velocity of propagation of the electromagnetic wave in the intense EM  background, and the aim of our paper is to determine the phase velocities that correspond to the birefringence as well as the electric field longitudinal component in terms of the magnetic background derived from the Euler-Kockel Lagrangian. Although these effects have been the subject of many studies, our approach is the framework  of the effective metrics  derived in nonlinear electrodynamics for curved spaces. 

The paper is organized as follows: In the next section the equations derived from the Euler-Kockel Lagrangian (\ref{E-HLagrangian}) are presented and the electromagnetic fields of the background and the wave are introduced; then by means of the effective metric approach we determine the birefringence and the two effective metrics, whose null geodesics are the light trajectories, determining then the corresponding phase velocities, the polarizations and the dispersion relations; in the same section we present the results when the background is a uniform electric field. In Section 3, by performing a Lorentz boost, we consider the background in movement, determining the velocity of the modeling medium and the changes that this situation introduces in the phase velocities of the propagation. Finally, conclusions are presented in the last section.

\section{The Euler-Heisenberg propagating wave}

In this section we derive the solutions to the EK field equations.
The nonlinear electrodynamics (NLED) field equations derived from the EK Lagrangian (\ref{E-HLagrangian}),
are

\begin{equation}
\label{Feqs}
 F_{\lambda \mu;\nu}+F_{\nu \lambda;\mu}+F_{\mu \nu;\lambda}=0; \qquad \partial_{\nu}\left[\sqrt{-g} \left( \mathcal{L}_{\rm F}F^{\mu\nu} + \mathcal{L}_{\rm G} F^{*\mu\nu}\right) \right]=0,
\end{equation}
where $\mathcal{L}_{\rm X}$ denotes the derivative of $\mathcal{L}$ with respect to the invariant $X$, $d \mathcal{L}/ dX$. 
\begin{figure}[H]
 \centering
 \includegraphics[width=0.6\textwidth]{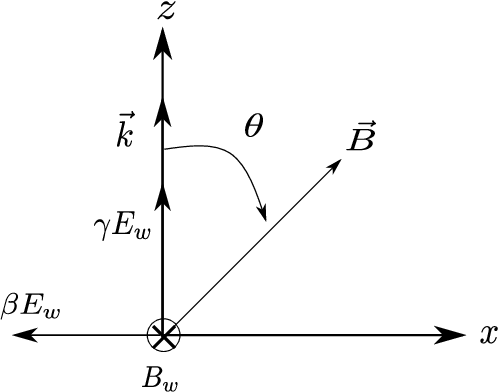}
 \caption{Plane wave electromagnetic fields with the electric component $\gamma E_w(\xi)$  arising  in the $\hat{z}$ direction, due to the nonlinear interaction. The uniform magnetic field background is shown making a $\theta$ angle with the $z$-axis.The magnitude of the fields is not on scale.}
 \label{fig1}
\end{figure}
Due to the interaction between the wave and the background,  
it is known that the parallel mode of the propagating wave fails to be orthogonal to the wave vector $\vec{k}$ \cite{Brezin1971}, because  
a longitudinal wave component arises in the presence of strong magnetic fields.

We shall consider an electromagnetic (EM) wave propagating  through a uniform magnetic background. The wave fields are function of  $\xi= (z- v t)$, where
$v$ is the phase velocity of the propagation in the $\hat{z}$ direction;
the electric and magnetic  fields of the propagating wave in terms of the electromagnetic potential $a^{\mu}$ are proposed as
 
 \begin{equation}
a_{\mu} (\xi) = (0, a(\xi),0, - \gamma a(\xi)/ v).    
 \end{equation}
We consider the arising electric component being $\gamma a(\xi)$, and the constant $\gamma$ is to be determined by solving the nonlinear electrodynamics (NLED) field Eqs. (\ref{Feqs}). The propagating vector $\vec{k}$ is along the z-direction, $k^{\mu}=(\omega,0,0,k)$, the wave electric field is along the $x$-direction and the wave magnetic field points to the $y$-direction.
 While for the uniform background  $A^{\mu}$
 
\begin{equation}
A_{\mu} (\xi) = (0, 0, B_x z- B_z x, B_y x)     
\end{equation}
 In Fig. \ref{fig1} is shown the plane wave propagating in the magnetic background; for convenience we locate the magnetic background in the plane XZ, $\vec{B} = B \sin \theta \hat{x} + B \cos \theta \hat{z} = B_x \hat{x} + B_z \hat{z}$.
 
 The photon polarization tensor in a homogeneous electromagnetic background in the context of QED has been addressed in \cite{Karbstein2013}, and
 in \cite{Aiello2007} was studied the case of a Born-Infeld wave in a magnetic background.
 In the next subsection we determine the phase velocities $v_i$, $i=1,2$ and $\gamma$ in the effective metric approach.

\subsection {Effective metric and  phase velocities of light rays}

It is well known that the intense EM fields can resemble a curved spacetime, in the sense that light trajectories are not straight lines  but  suffer deflection. Deviations from the straight trajectories in vacuum are described in NLED by the null trajectories of an effective metric.  According to this approach \cite{Pleban},  \cite{Novello2000},  \cite{Novello2000b}, the magnetic background distorts the spacetime where the EM wave propagates and considering the propagation as a perturbation, (i. e. the EM fields of the propagating wave are much smaller than the background fields) the effective metric is derived from the analysis of the propagation of the discontinuities of the field or characteristic surfaces (wavefronts). If $k_{\mu}$ is a null vector normal to the characteristic surface of the wave, the effective metric $g_{\rm eff}^{\mu\nu}$ is given by

\begin{equation}
\label{effgen}
 g_{\rm eff}^{(i) \mu\nu}k_{\mu}k_{\nu}=0, \quad i=1,2.
\end{equation}
The $(i)$ superscript corresponds to the two metrics that can arise in NLED, where the phenomenon of birefringence can occur. See \cite{Obukov2002} for a study on the Fresnel equation in nonlinear electrodynamics and \cite{Goulart2009} for a classification of the effective metrics. 

Eqs. (\ref{effgen}) are actually the dispersion relations, that for a wave number $k^{\mu}=(\omega,0,0,k)$ amount to

\begin{eqnarray}
\label{disp_rel}
&& g_{\rm eff}^{(i) tt} \omega^2 - 2 g_{\rm eff}^{(i) tz} \omega k  + g_{\rm eff}^{(i) zz} k^2 =0, \\
\label{disp_rel1}
\frac{\omega}{k}  & = & - \frac{g_{\rm eff}^{(i) tz}}{g_{\rm eff}^{(i) tt}} \pm \sqrt{ \left( \frac{g_{\rm eff}^{(i) tz}}{g_{\rm eff}^{(i) tt}}\right)^2 - \left({ \frac{g_{\rm eff}^{(i) zz}}{g_{\rm eff}^{(i) tt}}} \right) },  \quad i=1,2. 
\end{eqnarray}
The equations of the propagation of the field discontinuities in nonlinear electrodynamics characterized by a Lagrangian $\mathcal{L}(F,G)$  are given in  \cite{Novello2000b} [Eqs. (16) in \cite{Novello2000b} ]; in case of the EK Lagrangian, $\mathcal{L}_{FG}=0$ and  the effective metrics (\ref{effgen}) are given by,

\begin{eqnarray}
\label{eff1}
g_{\rm eff}^{(1) \mu\nu} &=& ( \mathcal{L}_{F} - 2 \mathcal{L}_{GG} F )\eta^{\mu\nu}-4\mathcal{L}_{GG}F^{\mu}{}_{\lambda}F^{\lambda\nu}, \\
\label{eff2}
g_{\rm eff}^{(2) \mu\nu} &=& \mathcal{L}_{F}\eta^{\mu\nu}-4\mathcal{L}_{FF}F^{\mu}{}_{\lambda}F^{\lambda\nu},
\end{eqnarray}
where $\eta^{\mu \nu} = {\rm diag} [+1,-1,-1,-1]$ is the Minkowski metric. In the Maxwell case  $\mathcal{L}=-F/4,\quad \mathcal{L}_{F}=-1/4, \quad \mathcal{L}_{FF}= 0 $ and $\mathcal{L}_{G}= 0$, then both effective metrics become  conformal to the Minkowski metric, $g_{\rm eff}^{(1) \mu\nu} = g_{\rm eff}^{(2) \mu\nu}= - \eta^{ \mu\nu}/4$, and the null geodesics coincide with the Minkowski ones. 
The nonvanishing electromagnetic tensor components of the magnetic background 
are $F^{xy}=-B_z, \quad F^{xz}= B_y= 0,\quad F^{yz}=-B_x,$ and $F=2 B^2$,
the corresponding phase velocities $v_i$ derived from the two effective metrics (\ref{eff1}) and (\ref{eff2}), through Eqs. (\ref{disp_rel}) are

\begin{eqnarray}
(v_{\rm  1})^2 &=& 1 - \frac{14 \mu B_x^2}{1 + 10 \mu B^2}, \\
(v_{\rm  2})^2 &=& 1 - \frac{8 \mu B_x^2}{1 -4 \mu B^2},
\label{vphi_from_eff_metr}
\end{eqnarray}

The effective metric approach turns out to be equivalent to a soft photon approximation.
Splitting the
total electromagnetic field into the background field $\vec{B}$ and the propagating photon $f_{\mu \nu}$, 

\begin{equation}
f_{\mu \nu} = \partial_{\mu} a_{\nu} -   \partial_{\nu} a_{\mu}=  (\epsilon_{\mu} k_{\nu} -   \epsilon_{\nu} k_{\mu}) e^{-ikx},  
\end{equation}
with the vector potential $a^{\mu} (k)$, the polarization $\epsilon_{\mu}= a_{\mu}/ \sqrt{a^{\alpha} a_{\alpha}}$, the wave vector $k^{\mu}= (\omega,0,0,k)$, and
being the phase $kx= k^{\alpha} x_{\alpha}= - k \xi$. Keeping the linear approximation with
respect to $f_{\mu \nu}$ in the equations of motion,  leads to an eigenvalue equation for the propagating
modes \cite{Liberati-Sonego-Visser}, \cite{Hu2007}

\begin{equation}
\label{amunu}
A^{\mu \nu} \epsilon_{\nu}=0    
\end{equation}
where $A^{\mu \nu}$ is given by

\begin{eqnarray}
\label{ecsA}
A^{\mu \nu} & = &  c_1 F^{\mu \alpha} F^{\nu \beta} k_{\alpha} k_{\beta} + c_2 F^{* \mu \alpha} F^{* \nu \beta} k_{\alpha} k_{\beta} +  c_3     ( \delta^{\mu \nu} \kappa^2 -  k^{\mu} k^{\nu}), \\
c_1 &=& \frac{1}{2} \mathcal{L}_{FF},\quad
c_2 = \frac{1}{2} \mathcal{L}_{GG}, \quad
c_3 = \frac{1}{2} \mathcal{L}_{F},
\end{eqnarray}
where $\kappa = k^{\alpha} k_{\alpha} = \omega^2-k^2, \quad k^{\mu}= (\omega, 0, 0, k)$.
Eq. (\ref{amunu}) is the light cone condition and its solutions are the dynamically allowed polarization modes.
Moreover, adopting the temporal gauge $\epsilon_0 = 0$, then Eq. (\ref{amunu}) splits into
\begin{equation}
A^{0i} \epsilon_0=0, \quad   A^{ij} \epsilon_j=0.  
\label{ecsAi}
\end{equation}

For the case under study, the nonvanishing electromagnetic tensor components of the magnetic background 
are $F^{xy}=-B_z, \quad F^{xz}= B_y= 0,\quad F^{yz}=-B_x,$ and their dual
$F^{* t x}=-B_x, \quad F^{* ty}= -B_y = 0, \quad F^{* tz}=-B_z$ and $F=2 B^2$.
Explicitly Eqs. (\ref{ecsAi}) become

\begin{eqnarray}
\label{ec_ep1}
&& c_2 B_z B_x \epsilon_1 + (c_2 B_z^2 -c_3) \epsilon_3  = 0, \\
\label{ec_ep3}
&& [c_2 \omega^2 B_x^2 - c_3 (\omega^2-k^2)] \epsilon_1 + c_2 \omega^2 B_z B_x \epsilon_3 = 0, \\
\label{ec_ep2}
&& [c_1 k^2 B_x^2 - c_3 (\omega^2-k^2)] \epsilon_2 =  0, 
\end{eqnarray}

The Eq. $A^{0i} \epsilon_0=0$ turns out to be the same than Eq. (\ref{ec_ep1}). Note that in Eqs. (\ref{ec_ep1}) and (\ref{ec_ep3})
are coupled $\epsilon_1$ and $\epsilon_3$; these Eqs. define the parallel polarization tensor as $\epsilon_{\parallel} = [0, a(\xi)/ka, 0, - \gamma a(\xi)/(\omega a)] $ or
$\epsilon_{\parallel} = [0, 1, 0, - \gamma/v] $ and determine 
the arising electric component $\gamma$ and a first dispersion relation as

\begin{eqnarray}
\label{gamma}
\gamma & = &\frac{14 \mu B_{x} B_{z}}{1-4\mu B^2+ 14 \mu B_z^2} v_{\rm 1} , \\
\label{v1}
\left(  \frac{\omega}{k} \right)_{(1)}^2  =  (v_{\rm 1})^2 &=&
1 - \frac{14 \mu B_x^2}{1 + 10 \mu B^2}.
\end{eqnarray}

The angle $\delta$ between the polarization $\epsilon_{\parallel}$  and the the propagating vector $\vec{k}$, is 
\begin{equation}
\delta = {\rm arccot} \left(\frac{\gamma}{v} \right)=   \frac{14 \mu B_{x} B_{z}}{1-4\mu B^2+ 14 \mu B_z^2}. 
\end{equation}

While Eq. (\ref{ec_ep2}) does not impose any condition on $\epsilon_2 \ne 0$, then
we can set the transversal polarization mode as $\epsilon_{\perp}=(0,0,1,0)$, and the second dispersion relation is

\begin{equation}
\left(  \frac{\omega}{k} \right)^2_{(2)} =   (v_{\rm 2})^2 = 1 - \frac{8 \mu B_x^2}{1 -4 \mu B^2},
\label{v2}
\end{equation}
where $B^2=B_x^2+B_z^2$. Transversal and parallel are defined with respect to the plane spanned by the magnetic field $B$ and the wave number $\vec{k}$, that is the plane $XZ$.
Phase velocities are in agreement with the ones derived from the effective metrics given by  (\ref{eff1}) and  (\ref{eff2}).

As a consequence of the nonlinear interaction wave-background there is a retarding term in $v_i$ that depends on the background field; specifically the retarding term arises due to the magnetic component that is perpendicular to the propagating direction, 
$B_x= B_{\perp}$, in such a way that if $B_{\perp} = 0$, then the velocity is the one in vacuum.

\begin{figure}[H]
\centering
\subfigure{
\includegraphics[width=0.49\textwidth]{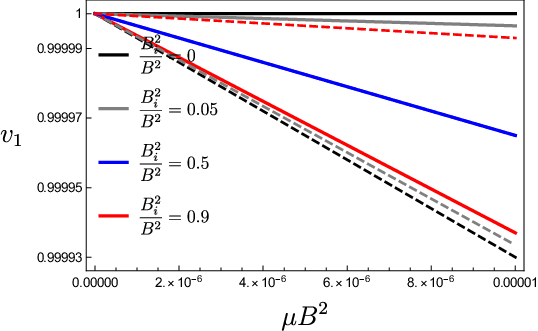}}
\subfigure{\includegraphics[width=0.49\textwidth]{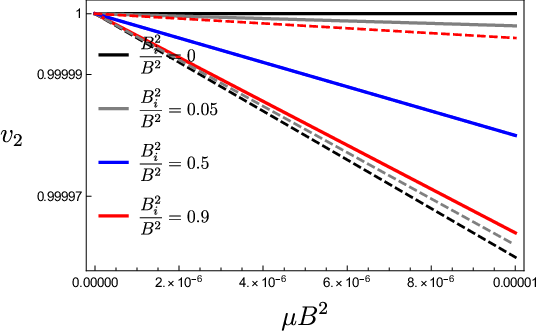}}
\caption{\small The phase velocities $v_{1}, \quad v_2$ are shown as a function of the (dimensionless) magnetic background $\mu B^2$, for four values of $(B_{x}/B)^2$ (thick lines) and $(B_{z}/B)^{2}$ (dashed lines).  As $\mu B^2$ increases $v_{1,2}$ diminish. If $\mu B^2 =0$ the velocity is the one in vacuum, $v_{i}=c=1$. If $(\frac{B_{x}}{B})^{2}=(\frac{B_{z}}{B})^{2}=\frac{1}{2}$, the corresponding velocities $v_{1}$ and $v_{2}$ behave the same  (blue line).
Note that the slowing effect increases as the transversal magnetic component $B_x$ does. From the behaviour of the thick and dashed lines we observe that the velocity is lower for values $(\frac{B_{x}}{B})^2<0.5$ and for values  $(\frac{B_{z}}{B})>0.5$.}
\label{fig2}
\end{figure}

\begin{figure}[H]
\centering
\includegraphics[width=0.6\textwidth]{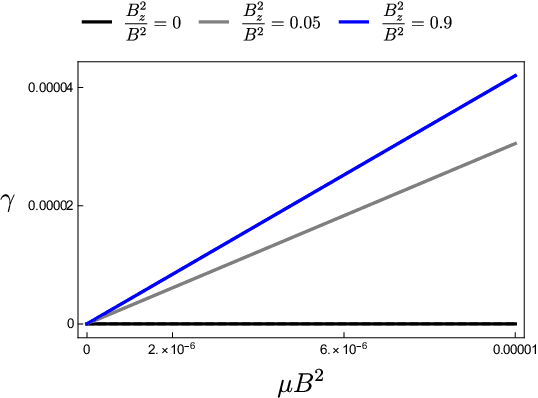}
\caption{\small  It is plotted $\gamma$, the arising wave electric component in the propagating direction, as a function of $\mu B^2$ for different values of $B_{z}^2/B^2$. }
\label{fig3}
\end{figure}

The phase velocities  $v_{\rm i}, \quad i=1,2 $ of the propagation through the magnetic background, Eqs.  (\ref{v1}), (\ref{v2}), are illustrated as a function of  the dimensionless $ \mu B^2$ in Fig.  \ref{fig2}.
The truncation we have used  to describe the propagation of photons in an external field is indeed valid for strong magnetic fields that may be as large as $10^{12}$ Gauss ( $10^{-1}$ Tesla), that compared  to the critic field  
$B_{\rm cr} \sim 10^{9}$ Tesla, gives us the range of
validity of our variable $\mu B ^2 \sim \alpha (B/B_{\rm cr})^2$, and considering that $B/B_{\rm cr} \sim 1/100$, the range of validity is  $ \mu B ^2 \in [0, 10^{-6})$. In agreement to this range the slowing down of the wave is on the hundred of thousandths. As shown in Fig. \ref{fig2} it would be very difficult to distinguish in an experiment if birefringence occurs.
In Fig. \ref{fig3} it is illustrated the factor $\gamma$ of
the arising wave electric component in the propagating direction.

In order that $v_{\rm 2}$ Eq. (\ref{v2}) be a real number, a lower bound arises for the magnetic field $B$,   $ \mu B^2 \le  1/4$, that in terms of the critical magnetic field $B_{cr}$ is $ \mu B^2 \approx \alpha (B/B_{cr})^2$, since $ (B/B_{cr}) << 1$ that value is never reached in this approximation.
If  $\mu=0$, that means the absence of vacuum polarization then the light velocity in vacuum, $v=c$, is recovered. 

\subsection{The electric uniform background}

In an analogous way to the previous subsection we  now consider an electric background.
The propagation of an EM wave through an intense uniform electric field
is of interest \cite{Adler1970} since there is the prediction of vacuum electron-positron production
that has not yet been measured, however it might be feasible in the near future, due to the high power reached lately by  lasers  \cite{Gies2009}, \cite{Karbstein2020}.

In this case
the nonvanishing electromagnetic tensor components for the electric background
are $F^{tx}=-E_x, \quad F^{ty}= -E_y,\quad F^{tz}=-E_z,$ and their dual
$F^{* xy}=E_z, \quad F^{* xz}= -E_y= 0, \quad F^{* yz}=E_x,$ and $F=-2 E^2= -2 (E_x^2+E_z^2)$.
The resulting equations from (\ref{ecsAi})  have the form of Eqs. (\ref{ec_ep1}), (\ref{ec_ep3}),(\ref{ec_ep2})  interchanging
$c_1 \leftrightarrow c_2$ and $B_i \leftrightarrow E_i$. Solving them we get the following $\gamma$-component and the phase velocities

\begin{eqnarray}
\gamma & = &\frac{8 \mu E_{x} E_{z}}{1 + 4\mu E^2+ 8 \mu E_z^2} v_{\rm  1} , \\
\left(  \frac{\omega}{k} \right)^2_{(1)}  =  (v_{\rm  1})^2 &=& 1 - \frac{14 \mu E_x^2}{1 + 4 \mu E^2}, \\
\left(  \frac{\omega}{k} \right)^2_{(2)} =   (v_{\rm  2})^2 &=& 1 - \frac{8 \mu E_x^2}{1 + 12 \mu E^2}.
\end{eqnarray}

\begin{figure}[H]
\centering
\subfigure{
\includegraphics[width=0.49\textwidth]{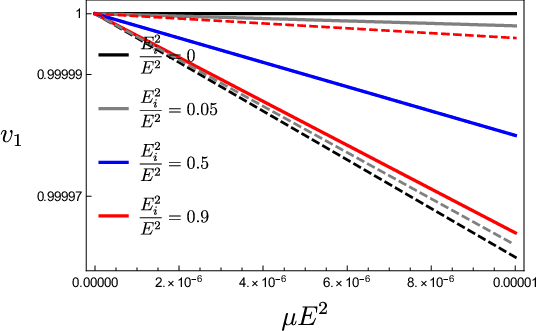}}
\subfigure{\includegraphics[width=0.49\textwidth]{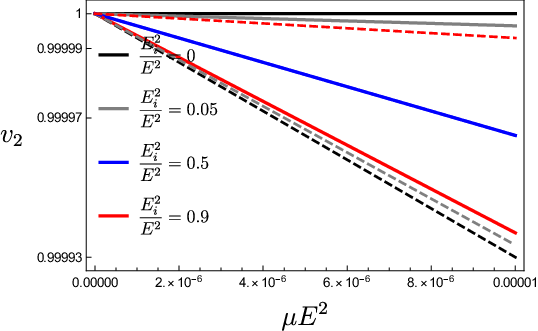}}
\caption{\small The phase velocities $v_{i}$ versus $\mu E^2$ are displayed. Velocities approach the ones in vacuum as $\mu E^{2} \mapsto 0$. Thick lines correspond to different $({E_{x}}/{E})^{2}$ and the dashed lines to  $({E_{z}}/{E})^{2}$. For the electric uniform background $v_2$ reaches lower values than $v_2$. When $({E_{x}}/{E})^{2}=({E_{z}}/{E})^{2}=\frac{1}{2}$ the velocities are the same   (thick line overlaps dashed line). }
\label{fig4}
\end{figure}

Phase velocities $v_1$ and $v_2$ can be derived as well from Eqs. (\ref{disp_rel}). $v_1$ and $v_2$ are shown in Fig. \ref{fig4}; the plots are very similar and for the shown scale there is not much qualitative difference. Analogous observations than for the magnetic background apply to the electric case.

\section{The phase velocity of the EM wave in a flowing magnetic background}

Another interesting situation is when the background is not still but moving with constant velocity; it can be considered as a plasma model, for instance.
Any effective metric can be considered as a moving medium, as we show in what follows, as long as the effective metric be nondiagonal. For the case we have studied this means that the metric component $g^{\rm{eff}}_{ tz}$ be nonvanishing. This we achieve by performing a Lorentz transformation on the NLED effective metric.

\subsection{The effective metric as a Painlev\'e-Lemaitre-Gullstrand (PLG) metric}

Let us consider a completely general effective metric (nondiagonal);
to determine the phase velocity of the EM wave $\beta =v/c$, we calculate the null geodesics of the effective metric, by making zero the line element, 
$ds^2= g^{\rm eff}_{\mu \nu} dx^{\mu} dx^{\nu} =0 $,

\begin{equation}
 \frac{ds^{2}}{d \tau^2}=g^{\rm{eff}}_{ \mu \nu} \frac{dx^{\mu}}{d \tau} \frac{dx^{\nu}}{d \tau}= g^{\rm{eff}}_{ \mu \nu} \dot{x^{\mu}} \dot{x^{\nu}} = 0. 
 \label{null_traj}
\end{equation}

Considering Cartesian coordinates $(t,x,y,z)$ and a light trajectory for fixed $x$ and $y$, ($\dot{x}=0= \dot{y}$), we obtain from Eq. (\ref{null_traj}) a quadratic equation for  the phase velocity  along the $z$-direction, ${\beta}= {dz}/{dt}= \dot{z} / \dot{t}$; then solving for $\beta = v/c$ we obtain the phase velocity in terms of the metric components of the effective metric, 

\begin{equation}
{\beta}=\frac{dz}{dt}= - \frac{g^{\rm{eff}}_{ tz}}{g^{\rm{eff}}_{ zz}} \pm  
\frac{\sqrt{ ({g^{\rm{eff}}_{ tz})^2-g^{\rm{eff}}_{ tt} g^{\rm{eff}}_{ zz}}}}{g^{\rm{eff}}_{ zz}}.
\label{beta}
\end{equation}
This expression is in agreement with Eq. (\ref{disp_rel1}).
The interpretation of the effective metric  as a propagating medium can be seen clearly by writing the effective metric $g^{\rm{eff}}_{\mu\nu}$ in the form of the
Painlev\'e-Lemaitre-Gullstrand (PLG) metric, that in Cartesian coordinates $(\tilde{t},\tilde{x},\tilde{y},\tilde{z})$ for $\tilde{x}=$const $ \tilde{y}=$ const, is given by
\begin{equation}
 ds^{2}=- (\tilde{c}^2 - V^2 ) d \tilde{t}^{2} -2V d \tilde{z} d \tilde{t} +d \tilde{z}^{2},
 \label{PLGmetricCart}
\end{equation}
where $V(\tilde{t}, \tilde{z})$ represents the velocity of the propagating medium and $ \tilde{c}(\tilde{t}, \tilde{z})$ the velocity of the perturbation propagating through such a medium \cite{Aguero2020}. Taking advantage of the constant curvature (scalar curvature is zero $R=0$) of the effective metric, by making a scale transformation on the $(t,z)$ coordinates we can write  the effective metric Eq. (\ref{null_traj}) in the PLG form (\ref{PLGmetricCart}). By re-scaling as

\begin{equation}
t \mapsto \frac{\sqrt{V^{2}-\beta^2}}{\sqrt{g_{tt}}} \tilde{t}, \quad z \mapsto \frac{1}{\sqrt{g_{zz}}} \tilde{z},  
\label{rescaling}
\end{equation}

the effective metric in the $(\tilde{t}, \tilde{z})$ coordinates acquires the form,

\begin{equation}
d\tilde{s}^{2}=-\left(\beta^{2}-V^{2}\right)d \tilde{t}^{2} +2g^{\rm{eff}}_{tz}\frac{\sqrt{V^{2}-\beta^{2}}}{g^{\rm eff}_{tt}g^{\rm eff}_{zz}} d \tilde{z} d \tilde{t}+d \tilde{z}^{2},
\label{PLGmetric}
\end{equation}
comparing with (\ref{PLGmetricCart}), we identify the velocity of the perturbation as $\tilde{c}=\beta$ and we determine the velocity of the medium $V$ as
\begin{equation}
V = \pm \frac{-g^{\rm eff}_{tz} \beta}{\sqrt{(g^{\rm eff}_{tz}{})^{2}-g^{\rm eff}_{tt}g^{\rm eff}_{zz}}}.
\label{V}
\end{equation}
Note that if $g^{\rm eff}_{tz} = 0$ then $V=0$, i.e. the medium is static.
By performing  a Lorentz boost of velocity $\beta_L$ in the $z-$direction $\Lambda_{z}$, on the effective metric, $\Lambda_{z} g_{\rm eff} \Lambda_{z}^T = g'_{\rm{eff}}$,  we obtain a  nondiagonal metric, that we denote with a prime $g'^{\rm{eff}}_{\mu\nu}$.
The effect of the Lorentz transformation  is of mixing the components of the effective metric, in such a way that if the effective metric is diagonal, the transformed one has nondiagonal components, i.e. $g'^{\rm{eff}}_{t z}$ does not vanish. The original (untransformed) effective metric is recovered when $\beta_L =0$.
Note as well that the velocity of the medium $V$ is not the same than the one of the Lorentz transformation $\beta_{L}$. The relationship between $V$ and $\beta_{L}$ is given implicitly  in Eq. (\ref{V}),
with $g^{\rm eff}_{\mu \nu} \mapsto  g'^{\rm{eff}}_{\mu\nu}$.

\subsection{The magnetic background Lorentz boosted}

In this subsection we show how the phase velocity is affected when a Lorentz boost is performed on the magnetic background.

Making a Lorentz transformation along the z-direction changes the magnetic background and an electric component field arises:
such that now the nonvanishing electromagnetic tensor components of the magnetic background are  $ F^{ty}= -\gamma_{L} \beta_{L} B_x,\quad F^{xy}=-B_z, \quad F^{xz}= B_y= 0,\quad F^{yz}=- \gamma_{L} B_x,$  and $F=2 B^2$.

Calculating the phase velocities using the effective metrics in Eqs.   (\ref{eff1}) and (\ref{eff2}), through Eq. (\ref{disp_rel1})
are

\begin{eqnarray}
v_{\rm  1}^{LTz} &=& \frac{-14 \mu \gamma_{L}^2 B_x^2 \beta_{L} \pm \sqrt{(1 + 10 \mu B^2)(1 + 10 \mu B^2 - 14 \mu B_x^2)}}{1 + 10 \mu B^2 + 14 \mu \gamma_{L}^2 \beta_{L}^2 B_x^2}, \\
v_{\rm  2}^{LTz} &=& \frac{-4 \mu \gamma_{L}^2 B_x^2 \beta_{L} \pm \sqrt{16 \gamma_{L}^4 \beta_{L}^2 \mu^2 B_x^4 + (1 -4 \mu B^2- 8 \mu \gamma_{L}^2 B_x^2 )( 1 -4 \mu B^2 + 8 \mu \beta_{L}^2 \gamma_{L}^2 B_x^2)}}{ 1 -4 \mu B^2 + 8 \mu \beta_{L}^2 \gamma_{L}^2 B_x^2},
\label{vphi_from_eff_metrLT-z}
\end{eqnarray}
where we are denoting $\beta_{L}$ as the velocity of the Lorentz transformation along $z$ and $\gamma_{L}^2 = (1- \beta_{L}^2)^{-2}$.
Although the factor $\gamma_{L} > 1$ enhances the retarding term, the denominator increases as well, in such a way that the effect of retarding does not increase much with the Lorentz transformation along the z-direction.

\begin{figure}[H]
\centering
\includegraphics[width=0.5\textwidth]{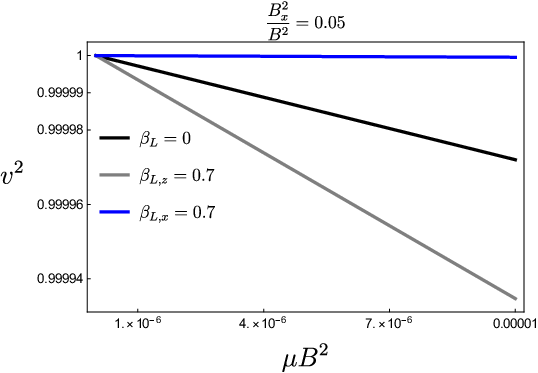}
\caption{\small  The phase velocities for metric 1 are shown in  three cases: $v_1$ for the metric Eq. (\ref{eff1}) (black) and then the corresponding to the Lorentz boost along the $z$ (grey) and $x$ (blue) directions. The slowing down is enhanced for the boost along the propagation direction. }
\label{fig5}
\end{figure}

Let us now try a Lorentz boost in the direction perpendicular to the propagation of the wave. By transforming the background with a Lorentz boost in the $x$-direction, the magnetic background results as  $ F^{ty}= \gamma_{L} \beta_{L} B_z,\quad F^{xy}=- \gamma_{L} B_z, \quad F^{xz}= B_y= 0,\quad F^{yz}=- B_x,$; in such a way that the phase velocities using the effective metrics in Eqs.   (\ref{eff1}) and (\ref{eff2}), through Eqs. (\ref{disp_rel1})
are

\begin{eqnarray}
v_{\rm  1}^{LTx}&=&  \frac{-14 \mu \gamma_{L} \beta_{L} B_x B_z \pm \sqrt{(1 + 10 \mu B^2)(1 + 10 \mu B^2 - 14 \mu( B_x^2-\gamma_{L}^2 \beta_{L}^2 B_z^2))}}{1 + 10 \mu B^2 + 14 \mu \gamma_{L}^2 \beta_{L}^2 B_z^2},\\
v_{\rm  2}^{LTx} &=&  \frac{8 \mu \gamma_{L} \beta_{L} B_x B_z \pm \sqrt{(1 -4 \mu B^2)(1-4 \mu B^2 +8 \mu( B_x^2-\gamma_{L}^2 \beta_{L}^2 B_z^2))}}{1 -4 \mu B^2 + 8  \mu \gamma_{L}^2 \beta_{L}^2 B_z^2},
\label{vphi_from_eff_metrLT-x}
\end{eqnarray}
where we are denoting $\beta_{L}$ as the velocity of the Lorentz transformation along $x$-direction and $\gamma_{L}^2 = (1- \beta_{L}^2)^{-2}$.
In Fig. \ref{fig5} are illustrated the velocities for metric 1 in three cases, $v_1$ corresponding to the metric Eq. (\ref{eff1}) and then the ones with the Lorentz boost along the $z$ and $x$ directions.
The direction of the Lorentz velocity is important: when $\beta_{L}$ is in the same direction than the propagation, the braking of the wave is more effective than if $\beta_{L}$ is perpendicular to the propagation. In the former case the wave slows down with a magnetic background less intense than for a still medium. 

It deserves a further study the modeling of a moving medium and specifically the precise relationship between the medium affects and the the velocity of the Lorentz tranformation. For the case of the electric background we guess has a similar behaviour than the magnetic one when Lorentz boosted.
\section{Conclusions}

We present the solutions to the truncated Euler-Heisenberg Lagrangian that represent an electromagnetic wave propagating through an intense uniform magnetic or electric background; due to the wave-background interaction one electric longitudinal component, $\gamma E_w$, arises affecting the polarization.. The constant $\gamma$ depends on three parameters:  the velocity of the propagation, the magnetic background and the EH parameter $\gamma (\beta, B, \mu)$; if the magnetic field is such that there is no component perpendicular or parallel to the propagating direction, $B_{\perp} =0$ or $B_{\parallel}=0$ this effect does not occur and $\gamma=0$.

By means of the NLED  effective metric approach \cite{Novello2000}, \cite{Pleban}, \cite{Liberati-Sonego-Visser}, we obtain 
there is birefringence and we determine the two possible phase velocities of the propagating wave. For  intense magnetic background fields,
but such that $B/B_{cr} << 1$ the velocities of the propagation  slow down, diminishing to the order of hundred thousandths as $B$ grows, this  is illustrated in Fig. \ref{fig2}. The birefringence in the case of an electric uniform background is calculated as well.

By performing a  Lorentz boost on the effective metric we model the situation of a flowing medium. By rescaling the coordinates of the effective metric it acquires the form of a  PLG metric, where clearly can be identified the velocity of the propagating medium and the velocity of the perturbation through such a medium. 
We present as example the case of Lorentz boosts  parallel and perpendicular to the propagating direction. The effect of slowing down the wave is more efficient for a parallel to the propagation Lorentz boost.

In summary we have analyzed the  slowing down of an electromagnetic propagating wave under the effect of a very intense electromagnetic field background, in the context of the truncated to one-loop Euler-Heisenberg theory, or the Euler-Kockel Lagrangian, that takes into account in an effective way the vacuum polarization phenomenon. We used the effective metric approach and showed it is equivalent to the soft photon approximation. We find there is birefringence in both cases the electric and the magnetic background, the phase velocities of the propagation depend on its polarization.

\vspace{0.5cm}
\textbf{Acknowledgments}: The work of E G-H has been sponsored by CONACYT-Mexico through the M. Sc. scholarship No. 485042.  N. B. acknowledges partial financial support from CONACYT-Mexico through the project No. 284489.

\end{document}